# Reflected wavefront manipulation by acoustic metasurface with anisotropic local resonant units

Meiyu Liu, Pan Li, and Pai Peng[a]

*School of Mathematics and Physics, China University of Geosciences- Wuhan, China*

**Abstract** –In this work, we develop the gradient metasurface is constructed of a locally anisotropic resonant structure, comprising a steel cylinder with an elliptical rubber coating embedded in epoxy. The deflective angles of rubber ellipses in the locally anisotropic resonant unit provide a method of controlling the reflected phase. Phase shifts of the reflected wave can cover the $2\pi$ range. With an appropriate design of the phase profiles along the acoustic metasurface, we can achieve anomalous reflection and Bessel beam. The locally anisotropic resonant units have significant potential for engineering and manipulating acoustic wavefronts.

Metamaterials are a class of subwavelength artificial composite structures with exotic properties that cannot be found in natural materials. However, these metamaterials have large thickness to lead to the bulky size and they go against miniaturization and integration of acoustic devices. A new type of metamaterials −metasurfaces − has emerged. Not only do acoustic metasurfaces exhibit the excellent ability of governing the acoustic wave, but also they provide a new method to accomplish the integration of acoustic devices. Inspired by optical wave manipulation, metasurfaces can be useful for manipulating acoustic wavefronts. The general Snell's law [1] is presented to rich the anomalous reflection and refraction phenomenon with phase discontinuities. Acoustic metasurfaces can freely tailor the wave field across a single layer and display many novel acoustic wave front control phenomena.

The acoustic metasurface is based on arrays of subwavelength units, including (but not limited to) Helmholtz resonators [2-6], membranes [7-11], coiling-up space [12-16] and locally anisotropic resonant structures [17]. Liu et al. [4] utilized Helmholtz resonators to theoretically demonstrate the apparent negative reflection when the incident angle is beyond the critical angle. Chen [11] also designed an acoustic composite structure of cavity and membrane to manipulate the phase profile in a controllable manner at low frequency. Li et al. [15] designed a coiling-up metasurface with discrete phase shift covering the full $2\pi$ span to realize the arbitrary adjustment of phase profiles. Liu et al. [17] designed a new acoustic metasurface based on locally anisotropic resonant units to realize the arbitrary control of the reflected wave in the water medium.

Similar the monopolar resonant mechanism of Helmholtz resonators, the locally anisotropic resonant units also have outstanding resonant performance. The three-component composites have a matrix of silicone-coated metallic spheres embedded in epoxy and were first proposed by Liu et al. [18]. They can generate a negative dynamic mass density at resonant frequencies. Then, the elliptical inclusions exhibit the anisotropic features. The elastic metamaterial built from lead cylinders with an elliptical rubber coating embedded in an epoxy matrix had an anisotropic effective mass density, based on a multi-displacement microstructure continuum model [19]. Following theoretical studies, microstructural designs with an anisotropic mass density have been experimentally validated [20].

In this work, we propose a resonant-based acoustic metasurface. Each of its anisotropic resonant units contains a steel cylinder with an elliptical rubber coating embedded in epoxy. It is an interesting phenomenon that the rubber ellipses in the locally anisotropic resonant unit provide another adjustable phase for the metasurface. The commercial software "COMSOL Multiphysics" (based on the Finite Element Method) was employed to numerically calculate. The dipolar resonant mechanism can be used to create adjustable phase shifts. Only one mode can be excited in the work frequency. These acoustic metasurfaces constructed by anisotropic resonant units with the deflective elliptic inclusion have an extraordinary reflected phase modulation and can be used to arbitrarily manipulate low-frequency acoustic waves in a water matrix.

[a]E-mail: paipeng@cug.edu.cn
China University of Geosciences - Wuhan China.

The anisotropic resonant units with the deflective elliptic inclusion contain the three accessible materials, which demonstrate remarkable phase change and novel acoustic property. Anomalous reflection and Bessel beam are exhibited by imposing suitable phase delay profiles of the metasufaces. The designed metasurfaces extend new degrees of freedom and paves the way for thin planar reflective structures for acoustic wave front manipulation.

Fig. 1(a) shows that when an acoustic wave is normally incident, the locally anisotropic resonant unit with deflective angles of the rubber ellipses reflects it completely. The dark purple and orange arrows indicate the propagation directions of the incident wave and reflected wave, respectively. Fig. 1(b) is a schematic of a locally anisotropic resonant unit. Each unit is composed of a steel cylinder with an elliptical rubber coating embedded in epoxy. The deflective angles of the rubber ellipses can be adjusted. The parameters of the materials used are: $\rho = 1000 kg/m^3$, $c = 1490 m/s$ for water; $\rho_e = 1180 kg/m^3$, $\lambda_e = 4.4 \times 10^9 N/m^2$ and $\mu_e = 1.6 \times 10^9 N/m^2$ for epoxy; $\rho_r = 980 kg/m^3$, $\lambda_r = 1.6 \times 10^9 N/m^2$ and $\mu_e = 5.5 \times 10^5 N/m^2$ for rubber; $\rho_s = 7900 kg/m^3$, $\lambda_r = 1 \times 10^{11} N/m^2$ and $\mu_e = 8.1 \times 10^{10} N/m^2$ for steel, where $\rho$ is the mass density, $c$ is the speed of sound, and $\lambda$ and $\mu$ are the Lame constants.

We designed and optimized 16 types of anisotropic resonant units with trihedral rigidness, to produce phase delay profiles with a relatively high resolution. The units have a square cross section of length w. The semimajor and semi-minor axes of the rubber ellipses are fixed as 0.3w and 0.25w, respectively. The reflected phase can cover $2\pi$ phase span along with the gradually deflective angles of the rubber ellipses. In the planar acoustic metasurfaces, for simplicity, the radii of the steel cylinders are tailored to yield the desired phase shifts. The phase of the reflected waves as a function of deflective angles of the rubber ellipses is plotted in Fig. 1(c). The three selected vibration modes are shown in Fig. 1(c). The wavelength of incident acoustic wave is 8.8w, which is constant in the following text. Note that the 16 units designed can realize a phase span of $2\pi$ with an interval of $\pi/8$ between adjacent units, as denoted by the red dots. By applying the condition of prescribed velocity, it is possible to keep adjacent units independent of each other. When appropriate reflected phases are selected, each unit can be simulated individually. Thus, these locally anisotropic resonant units have potential as building blocks for wavefront-shaping metasurfaces.

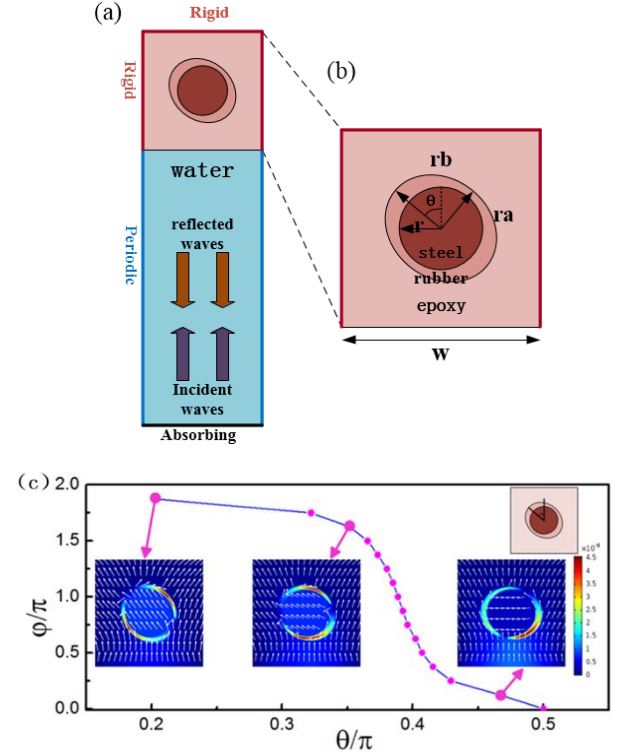

Fig. 1: (a) Schematic diagram of the numerical simulation of the reflective coefficient of the locally anisotropic resonant unit with the deflective elliptic inclusion. The dark purple and orange arrows refer to the propagation directions of the incident wave and reflected wave, respectively. The left and right boundaries (marked by blue lines) of water have been defined as the periodic boundary conditions and the bottom (marked by black line) is the absorbing boundary. The three boundaries of the anisotropic resonant unit (marked by red lines) have been set as acoustic rigid boundaries. (b) Schematic of the proposed anisotropic resonant unit, which comprises a steel cylinder with an elliptical rubber coating embedded in epoxy. The deflective angles of the rubber ellipses can be adjusted, whereas the semi-major and semi-minor axes of the rubber ellipses are fixed as $r_b = 0.3w$ and $r_a = 0.25w$, respectively. The units have a square cross section of length w. The deflective angles of the rubber ellipse θ are set

as an adjustable parameter. (c) Phase of the reflected wave as a function of deflective angles of the rubber ellipses for incident wavelength λ = 8.8w.

To design an acoustic metasurface with nearly arbitrary wave front modulation and planar profile, some locally anisotropic resonant units should be selected under the guidance of the general Snell's law. The general Snell's law can predict the anomalous propagation of the incident acoustic wave. Both the reflected wave and refracted wave are suitable. The direction of abnormal reflection is related to the direction of incident acoustic waves. The formula is as follows [1]:

$$\sin(\theta_r) - \sin(\theta_i) = \frac{d\varphi(x)}{k_0 dx} \quad (1)$$

Where $\theta_r$, $\theta_i$ are the reflected and incident angles, respectively; $\varphi(x)$, $dx$ are the phase discontinuities and the distant between the crossing points along the x direction, respectively; $k_0 = 2\pi/\lambda$ is the wave vector in water. Equation (1) shows the direction of reflected wave can be controlled freely through a suitable gradient phase profile.

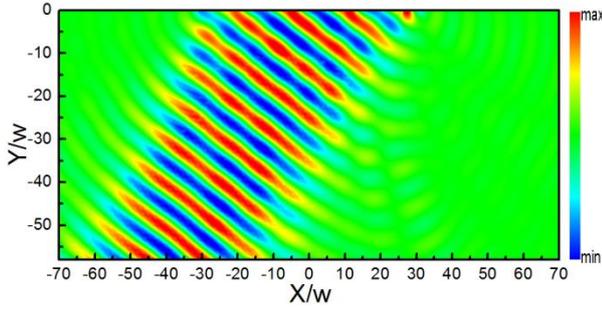

Fig. 2: The wavelength of incident acoustic wave is 8.8w. Acoustic reflected pressure field pattern for gradient phase profile is shown, when the acoustic wave impinge the metasurface from the bottom to the top. The phase gradient of the designed acoustic metasurface is set as $d\varphi(x)/dx = \pi/8w$.

The numerical simulation of acoustic metasurfaces based on the deflective elliptic inclusion units is introduced. When an acoustic wave from the bottom is normally incident toward the designed metasurfaces, the reflected pressure fields show that the acoustic wave is oblique. The angle of reflected wave is calculated as $33.4°$ utilizing the general Snell's law. From the reflected pressure distribution, it is noted that the simulated results agree well with the theoretical ones. And the locally anisotropic resonant units with the deflective elliptic inclusion are also capable of controlling reflected acoustic waves arbitrarily and open up a new avenue for acoustic wave front engineering and manipulations.

In the following, we show the different wave manipulation effect based on the different combinations of fundamental units. A planar acoustic axcion is exhibited. In order to the design the planar acoustic axcion, the reflected angle of sound wave is set as 10. When an acoustic wave from the bottom is normally incident, the reflected phase gradient of the metasurface along any x position should satisfy the next equation:

$$\varphi(x) = k_0|x|\sin\beta \quad (2)$$

Fig. 3: (a) A desire hyperboloidal phase profile along x direction of reflected wave for the designed lens. (b) The transverse distribution of acoustic pressure intensity profile at y=20.5m. (c) The pressure field distribution of reflected wave for the designed metasurface with $f = 3\lambda$, as the incident acoustic waves propagate from the bottom.

Applying the aforementioned condition, when the focal length is set as $3\lambda$, the structure of three-component composites along any x position is certain. Fig. 3(a) illustrates the continuous phase shifts along the x axis of reflected wave. The spatial reflected pressure intensity distribution is shown in Fig. 3(c). To better exhibit the performance of the acoustic lens, the transverse cross-section pressure intensity distribution at y=20.5 m is plotted in Fig. 3(b).

In this work, we present an acoustic metasurface based on anisotropic resonant units with deflective angles of the rubber ellipses, which have sufficient degrees of freedom for wavefront modulation. By selecting an appropriate gradient phase profile, the acoustic metasurface can realize anomalous reflection. It illustrates the remarkable ability of controlling the acoustic wave. The acoustic metasurfaces built from our anisotropic resonant units are a new approach for engineering and manipulating wavefronts.

.

***

This work was supported by the National Natural Science Foundation of China (Grant No: 11604307)